\documentclass[11pt]{article}

\usepackage{cite}
\usepackage{amsfonts}
\usepackage{amssymb}
\usepackage{amsmath}
\usepackage{bm}

\def\CC{\mathbb C}
\def\EE{\mathbb E}
\def\FF{\mathbb F}
\def\HH{\mathbb H}
\def\RR{\mathbb R}
\def\ZZ{\mathbb Z}
\def\too{\rightarrow}

\usepackage{latexsym}
\usepackage{mathrsfs}
\usepackage{a4}
\usepackage[dvips]{rotating}
\DeclareMathAlphabet{\mathpzc}{OT1}{pzc}{m}{it}
\usepackage[all]{xy}
\usepackage{makeidx}
\usepackage{graphicx}
\usepackage{pstricks}
\usepackage{multicol}
\usepackage{fancybox}
\input xy.tex
\xyoption{all}
\providecommand{\U}[1]{\protect\rule{.1in}{.1in}}
\newtheorem{theorem}{Theorem}

\newtheorem{lemma}[theorem]{Lemma}

\newtheorem{proposition}[theorem]{Proposition}

\begin{document}

\title{{Clifford Algebras in Symplectic Geometry and Quantum Mechanics}}
\author{Ernst Binz$^1$,  Maurice A. de Gosson$^2$, Basil J. Hiley$^3$.\\
\small{$^1$ Fakult\"{a}t f. Mathematik und Informatik A5/6, D-68131 Mannheim, Germany.}\\
\small{$^2$ Universit\"{a}t Wien, NuHAG, Fakult\"{a}t f\"{u}r Matematik, A-1090 Wien.}\\
\small{ $^3$ TPRU, Birkbeck, University of London, London, WC1E 7HX.}}
\date{\small{Accepted. To appear in Fond. Phys.\\arXiv:1112.2378 (math-ph)\\DOI:1007/s10701-012-9634-z}}

\maketitle

\begin{abstract}

The necessary appearance of Clifford algebras in the quantum description of fermions has prompted us to re-examine the fundamental role played by the quaternion Clifford algebra, $C_{0,2}$.  This algebra is essentially the geometric algebra describing the rotational properties of space.  Hidden within this algebra are symplectic structures with  Heisenberg algebras at their core.  This algebra also enables us to define a Poisson algebra of all homogeneous quadratic polynomials on a two-dimensional sub-space, $\FF^a$ of the Euclidean three-space.  This enables us to construct a Poisson Clifford algebra, $\HH_F$, of a finite dimensional phase space which will carry the dynamics.  The quantum dynamics appears as a realisation of  $\HH_F$ in terms of a Clifford algebra consisting of Hermitian operators.
\end{abstract}

\section{Introduction}

In a series of lectures at the annual Askloster Seminars, hosted by Georg Wikman, one of us (BJH) developed a new
approach to quantum mechanics based on the notion of `structure process', which provides a very different way
of looking at quantum phenomena \cite{bjh_2011process}. There it was shown how orthogonal Clifford algebras
arise from a collection of what we call `elementary processes'.  It should be no surprise that Clifford algebras
play an essential role in quantum mechanics since these algebras are necessary for a description of both the
Pauli and Dirac particles.  The fact that the Schr\"{o}dinger particle can also be described in terms of a
Clifford algebra may come as a surprise.

In this paper we will not discuss these elementary processes in their full generality.  Rather we will take
these ideas to motivate a discussion of the role of quaternions in bringing quantum mechanics and geometry
closer together.  Another motivation for our present discussion comes directly from the work of Binz and Pods
\cite{Binz_Pods_2008}.  This shows that a natural symplectic structure can be constructed using the quaternions.
This, in turn, gives rise to the Heisenberg algebra and it is the appearance of this algebra as a geometric
feature that enables us to make contact with quantum mechanics.

We start by constructing the skew-field of quaternions, a Clifford algebra.  We call this algebra, together with
the Clifford algebra of an Euclidean plane, the elementary Clifford algebra.  Combining this structure with the
complex numbers  (which drop out of the real quaternions), the general theory of real Clifford algebras as
presented in Greub \cite{Greub_1978}, then allows us to construct the (finite-dimensional) Clifford algebras of
bilinear forms of arbitrary index out of the elementary ones, thus enabling us to discuss the Schr\"{o}dinger,
Pauli and Dirac particles in one geometric structure.

In quantum mechanics the symplectic geometry plays a fundamental role through the Heisenberg group as has been
shown by de Gosson \cite{deGosson_M}. How is this structure hidden in Clifford algebras? To answer this question
we will focus, in particular, on the geometric aspects of the quaternion Clifford algebra  which encodes a
two-dimensional \emph{symplectic space} $\Bbb F^a$, characterised by selecting a particular  pure quaternion
$a$. $\FF^a$ is the tangent plane to the two-sphere at $a$.  The choice of $a$ is arbitrary.  We could have chosen
another pure quaternion, say $b$, and this will give rise to a different two-dimensional symplectic space
$\Bbb F^b$.  However these pure quaternions are related by the Clifford group $SU(2)$, so that we have an infinity
of Heisenberg algebras within the quaternion Clifford, $\HH=\RR\cdot {\bm e} \oplus \EE$, with $\EE$ being the
 Euclidean three-space.

Each symplectic space gives rise to a two-dimensional classical \emph{phase space},  while the whole structure
gives rise to a $2n$-dimensional phase space, which in turn, gives rise to a Poisson algebra.  We then find the
Poisson algebra of all homogenous quadratic polynomials over a two-dimensional \emph{phase space} determines a
natural Clifford algebra as well, which we will call the \emph{Poisson Clifford algebra} of the phase space.
This is isomorphic to the geometric Clifford algebra. The isomorphism is based on the construction of
Hamiltonian vector fields. The Poisson algebra of homogenous quadratic polynomials of any phase space of
dimension higher than two is  the linear image of a natural subspace of the Poisson Clifford algebra. Finally we
introduce a realization of the characteristic Clifford algebra of a phase space generated by Hermitian operators.
This realization is based on the quantization map of the above mentioned Poisson algebra. This final step is
where we make contact with quantum mechanics.  This elementary presentation follows essentially the book ``The
Geometry of Heisenberg Groups" \cite{Binz_Pods_2008}.

\section{From Processes to Clifford algebras}

\subsection{Processes}\label{PrH}

In Hiley \cite{bjh_2011process} the notion of  an elementary process was introduced.  A set of these elementary
processes can then be ordered by defining a simple multiplication rule, which then produces a structure that is
isomorphic to the quaternion Clifford algebra. A detailed discussion of the physical motivation and the underlying
philosophical grounds for introducing the fundamental notion of a \emph{process} and its history, we refer to
Hiley \cite{bjh_2011process}, Kauffman\cite{kauffman1991knots} as well as to Penrose \cite{penrose2004road}.

The fundamental idea behind this approach is to show how a geometric picture of an Euclidean three-space, $\EE$,
emerges from the order of the elementary processes encapsulated in the algebra.  We can represent each
\emph{process} as  a \emph{directed} one-simplex $[P_{1}P_{2}]$, connecting the opposite distinguishable poles
of the process $P_1$ and $P_2$.  As a basic assumption, we regard these process as {\em indivisible}, meaning
that the poles are merely distinguishing features of the process that cannot be separated from the process
itself. Since we are assuming our processes are directed, we assume $[P_{1}P_{2}]=-[P_{2}P_{1}]$.  The totality
of one-simplexes forms a real linear space isomorphic to the rotational symmetries of $\EE$.

An \emph{inner multiplication} for the one-simplexes is defined by
\begin{eqnarray}\label{1.1.pr}
[P_{1}P_{2}]\bullet[P_{2}P_{3}]=[P_{1}P_{3}].
\end{eqnarray}
This multiplication rule encapsulates the Leibnizian notion of the order of succession.

Starting with the three elementary processes $[P_0 P_1]$, $[P_0 P_2]$ and $[P_1 P_2]$ we find

\begin{eqnarray}\label{1.1.0}
\begin{tabular}{c|ccc}
1&$\;\;[P_0 P_1]$&$\;\;\;[P_0 P_2]$&$\;\;\;[P_1 P_2]$\\ \hline
$[P_0 P_1]$&$-\bm e$&$-[P_1 P_2]$&$\;\;\;[P_0 P_2]$\\
$[P_0 P_2]$&$\;\;\;[P_1 P_2]$&$-\bm e$&$-[P_0 P_1]$\\
$[P_1 P_2]$&$-[P_0 P_2]$&$\;\;\;[P_0 P_1]$&$-\bm e$
\end{tabular}
\end{eqnarray}
In this structure $[P_0 P_0]$, $[P_1 P_1]$ and $[P_2 P_2]$ are easily shown to act as two sided unities,
which we identify and denote by a single unity $\bm e$.  In this way we have generated a group from the
elementary processes. We can quickly identify this group by making the following notation change
$\bm i:=[P_0,P_1]$, $\bm j:=[P_0,P_2]$ and $\bm k:=[P_1,P_2]$. In this way we immediately see that
$\bm e,\bm i,\bm j,\bm k$ generate the quaternion group.  The resulting skew-field $\HH$ of
\emph{quaternions} is a real \emph{Clifford algebra} \cite{Greub_1978},
which can be verified by the defining universal property presented below.
The quaternions are often called a hypercomplex system.

In \cite{bjh_2011process} it was shown how to generalise the process algebra to include the
Pauli- and Dirac-Clifford algebras.  This has enabled Hiley and Callaghan \cite{bjhrc11C}
\cite{hiley2010cliffordA}\cite{hiley2010cliffordB} to apply these algebras in a novel way
to the quantum mechanics of the Schr\"{o}dinger, Pauli and Dirac particles.  This  has led
to a systematic generalisation of the Bohm approach \cite{Bohm_Hiley_1993} to the relativistic domain.

Our approach should not be confused with the attempt made by Finkelstein {\em et al.}
\cite{Finkelstein:1962fk} to develop a quaternion quantum mechanics.  These authors were
investigating the possibility of using a Hilbert space over the quaternion field.
Our approach is exploring a very different structure.

\subsection{Clifford Algebras}\label{CA}

For convenience we repeat the notion of a Clifford algebra by means of its \emph{universal property}. Let $\cal A$ be
an associative algebra with a unit element $e_A$. A \emph{Clifford map} from the $\RR$ -linear space $\FF$ to $\cal A$
is a linear map $\varphi$ which satisfies $(\varphi(h))^2=b(h,h)e_A$. Here $b$ is an $\RR$-valued, \emph{symmetric
bilinear} map defined on $\FF$. The \emph{universal property} mentioned above reads as follows:

A Clifford algebra $C_F$ over $\FF$ is an associative algebra $C_F$ with unit element $e$, together with a
\emph{Clifford map} $i_F$ from $\FF$ to $C_F$ subject to the following two conditions: $i_F(F)\subset C_FF$
generates $C_F$ and to every Clifford map $\varphi$ from $\FF$ to $C_F$, there is a unique \emph{algebra
homomorphism} $\Phi$ from $C_F$ to $\cal A$ satisfying $\varphi=\Phi\circ i_F$.

$\HH$ is a very basic Clifford algebra as we will see in the next section. The bilinear map is $-< , >$, the
negative of a scalar product $< , >$ on a two-dimensional subspace $\FF$ in $\EE$ (cf.~\cite{Greub_1978}).

Let us point out at this stage that the $\RR$-algebra End\,$F$ with the \emph{composition} as its product is a
Clifford algebra. It is determined by a scalar product $< , >$ on $\FF$ (cf.~\cite{Greub_1978}). The
multiplication table for $J,A,B\in \mbox{End} F$ with respective matrices
\begin{eqnarray} \label{MTEND}
J=\begin{pmatrix}0&1\\-1&0\end{pmatrix};\qquad
A=\begin{pmatrix}1&0\\0&-1\end{pmatrix};\qquad
B=\begin{pmatrix}0&1\\1&0\end{pmatrix}
\end{eqnarray}
is
\begin{eqnarray}\label{MT End F}
\begin{tabular}{c|ccc}
id&J&A&B\\ \hline
J&-id&-B&A\\
A&B&id&J\\
B&-A&-J&id
\end{tabular}
\end{eqnarray}
This is due to
\begin{eqnarray}\label{MTEND 2}
A\circ J=B,\qquad J\circ B=A,\qquad A\circ B=J,\qquad\text{ and }\qquad [A,B]=2\cdot J
\end{eqnarray}
as a direct calculation shows. (See Binz and Pods \cite{Binz_Pods_2008}). Here $[A,B]$ is the \emph{commutator} of $A$ and $B$.

This multiplication table (\ref{MT End F}) differs only in the signs down the diagonal from table (\ref{1.1.0}) and so can be regarded as arising from the structure process in $\FF$.  In this case they are isomorphic to the Clifford algebra End\,$F$.

In fact, any real finite-dimensional Clifford algebra of a symmetric bilinear form of arbitrary index is a
finite graded tensor product of the Clifford algebras $\CC$, $\HH$ and End\,$F$ (cf.~\cite{Greub_1978}). In this
sense the processes introduced above determine all these Clifford algebras.

We leave the algebraic treatment of processes and focus on another concept within real Clifford algebras. We will show how these algebras contain symplectic spaces, a basic entity of Hamiltonian mechanics and  quantum mechanics. Of course, these two approaches to Clifford algebras hint at a relation between processes that could give rise to symplectic geometry, to Hamiltonian mechanics and to quantum mechanics. The latter relations are discussed in Hiley \cite{bjh_2011process}.

\section{Clifford Algebras of Symplectic Spaces}

\subsection{The geometric Clifford Algebra of a symplectic space}\label{sbsec1}

In this subsection we will show how a characteristic Clifford algebra emerges from each (\emph{2n-dimensional})
symplectic space.

To this end we start with $\FF$ considered as  a \emph{two-dimensional}, real liner space equipped with a (constant)
non-degenerated, skew-symmetric bilinear two-form $\omega$, a \emph{symplectic form}. Hence $\FF$ is a
two-dimensional \emph{symplectic space} (cf.~\cite{deGosson_M}). In two dimensions, $\omega$ serves a
\emph{volume form} on $\FF$. In other words $\FF$ is an \emph{oriented}, two-dimensional linear space
(cf.~\cite{deGosson_M}).

Let us analyse this assumption in a little more detail. For simplicity we choose a scalar product $< , >$ on $\FF$ and represent
$\omega$ by a \emph{skew-adjoint} isomorphism $S$ so that

\begin{eqnarray}\label{1.1.1}
\omega (v,w)=<S(v),w>\qquad\forall u,v\in \FF.
\end{eqnarray}
There is an orthonormal basis in the Euclidean space, $\FF$, such that $S$ has the form
$$
S=\begin{pmatrix}\;\;0&\kappa\\-\kappa&0\end{pmatrix};\quad\quad \kappa\in \RR.
$$
Without loss of generality, we may choose $\kappa=1$ if $< , >$ is rescaled
suitably; in this case we write $J$ instead of $S$ and its matrix is
\begin{eqnarray}\label{acs}
J=\begin{pmatrix}\;\;0&1\\-1&0\end{pmatrix}\hspace{1.6cm}
\end{eqnarray}
as represented in (\ref{MTEND}). This map gives rise to an \emph{almost complex structure}.

Therefore, we conclude that, given a symplectic space $\FF$ with skew form $\omega$, there is some scalar product
and an isomorphism $J$ satisfying
\begin{eqnarray}\label{1.1.2}
J^2=-id \label{first}
\end{eqnarray}
representing $\omega$ in the sense of (\ref{1.1.1}).

Vice versa, given $\omega$ and some $J$ fulfilling (\ref{1.1.2}), there is a \emph{unique} scalar product $<
,>$, given by
\begin{eqnarray}\label{1.1.3}
-<v,w>:=\omega (J(v),w)\qquad\forall u,v\in \FF.
\end{eqnarray}

In fact, $J\in Sp(F)$, the group of all isomorphisms of $\FF$ preserving the symplectic form. This group is called the \emph{symplectic group}.

To construct a \emph {phase space} out of $\FF$ we have to specify a \emph{configuration space}, i.e. the space of
positions in $\FF$. This space is a real line $\RR \cdot e_1 \subset \FF$ spanned by the vector $e_1 \in \FF$, say.
Then the line $\RR\cdot J(e_1) \subset \FF$ represents the space of \emph {momenta}. $\FF$ decomposed in this way is
a \emph{phase space}. For simplicity of notation let $e_2:=J(e_1)$. The basis $\{e_1, e_2\}\subset \FF$ can be assumed
to be symplectic, i.e. $\omega(e_1,e_2)=1$.  We will denote the coordinates by $q$ and $p$, respectively. The symplectic
form $\omega$ on the pase space $\FF$ is then the \emph{canonical} one (cf.~\cite{Marsden_Ratiu_1994}). Hence

\begin{lemma}\label{splitF}
Given a symplectic space $\FF$ with symplectic form $\omega$ and some almost complex structure $J \in  Sp(F)$ with
$J^2=-id$, there is a unique scalar product $<,>$ defined by
\[
<v,w>=-\omega (J(v),w)\qquad\forall u,v\in\FF.
\]
Moreover, given any unit vector $e_1\in \FF$ the basis $\{e_1,e_2\}$ with $e_2:=J(e_1)$ is a symplectic and
orthonormal one. This is to say it spans the phase space
\[
\FF=\RR\cdot e_1\oplus\RR\cdot e_2.
\]

\end{lemma}
Let us remark that associated with $J$, the symplectic space $\FF$ is turned into a \emph{complex line}. To see
this, we form the two-dimensional $\RR$-linear space
\[
\CC^J:=\mbox{span}\{id,J\}.
\]
Here $id, J \in Sp(F)$. The operations on $\CC^J$ will be defined in a similar way to those on a  field of complex variables; $J$ plays the role of the complex unity $i$ in $\CC$. The action of the field $\CC^J$ on $\FF$
is determined by the action of the linear maps $id$ and $J$ on $\FF$. This turns $\FF$ into a complex line.

The symplectic space $\FF$ can naturally be enlarged to a three-dimensional, oriented, Euclidean, real linear
space
\begin{eqnarray}
\EE:=\FF\oplus \RR\cdot J.
\end{eqnarray}
We extend $<,>$ orthogonally to a scalar product $<,>_E$ on all of $\EE$ for which $<J,J>_{E}=1$  and fix a metric
volume form $\mu_{E}$ on $\EE$. If its restriction to $\FF$ is $\omega$, then
\begin{eqnarray}\label{1.1.4}
\mu_{E}(e_1,e_2,J)=1.
\end{eqnarray}
The oriented, Euclidean linear space $\EE$ then admits a \emph{cross product}. With these ingredients we turn
the real linear space
\begin{eqnarray}
\HH:=\FF\oplus \CC^J=\EE\oplus \RR\cdot id
\end{eqnarray}
into a skew-field by setting
\begin{eqnarray}\label{quat}
(\lambda_1 e+u_1) \cdot(\lambda_2 e+u_2):=(\lambda_1\lambda_2-<u_1,u_2>)\cdot e +\lambda_1 u_2+\lambda_2 u_1+u_1
\times u_2
\end{eqnarray}

for all $\lambda_1, \lambda_2 \in \RR$ and any $u_1, u_2 \in \EE$.  Here $e$ replaces $id$. This skew-field is
called the \emph{quaternions} $\HH_{F}$ (cf.~\cite{Greub_1975}).

From the matrix structure of $J$, we deduce
\begin{eqnarray}\label{1.1.5}
J(v)= j\times v = j\cdot v\quad \forall v \in \FF.
\end{eqnarray}
for a \emph{unique} unit vector $j\in \EE$, perpendicular to $\FF$; hence $\FF$ is the tangent plane to the
unit sphere $S^2$ in $\EE$ at $j$. As a consequence $j\cdot j=-e$. Because of (\ref{1.1.5}), we identify the
vector $j\in \EE$ with the map $J$ on $\FF$. Hence, in dealing with $\HH$ we will replace $J$ by $j$ in what
follows.

By the universal property, $\HH_{F}$ is the \emph{Clifford algebra} of the plane $\FF$ endowed with $-<,>$
(cf.~\cite{Greub_1978}) and hence we call it the \emph{geometric Clifford algebra} of the symplectic space $\FF$.  Occasionally, we express this by the symbol $\FF^j$.
Vice versa, by (\ref{quat}) the algebra $\HH_{F}$ determines the bilinear form $-<,>$ on $\FF$, because
$h^2=-<h,h>$ holds for any $h\in \FF$. Moreover, $C^j$ is a commutative subfield of $\HH_{F}$ and complements
the linear space $\FF$ within $\HH_{F}$ and thus the Clifford algebra $\HH_{F}$ is ${\ZZ}_2$ graded.

The space $\FF$ is embedded as a phase space if an orthonormal basis ${h,k}\subset \FF$ is chosen. In this case we
have the orthonormal basis ${h,k,j,e}$ in $\HH_{F}$.

The unit sphere $S^3\subset\HH_{F}$ is a group, namely $SU(2)$ (cf.~\cite{Benn_Tucker_1987}, of which the
two-sphere $S^2$ in $\EE$ is the equator. $S^2$ constitutes all $j\in \EE$ with $j^2=-e$. Alternately formulated, it constitutes all the almost complex structures $J$ on all two-dimensional linear
subspaces in $\EE$.

Clearly $\omega$ is algebraically encoded in the Clifford algebra $\HH_{F}$ by (\ref{1.1.5}) and
(\ref{1.1.1}), or more sophisticatedly expressed, the Lie bracket of the \emph{Heisenberg algebra} on $\EE=\FF\oplus \RR \cdot j$ given by $\omega$ is determined by the product in $\HH_{F}$
(cf.~\cite{Binz_Pods_2008}). The metric $-<,>_E$, being part of the product in $\HH_{F}$, plays a central role
in special relativity (cf.~\cite{Binz_Pods_2008}). More generally, $\HH_{F}$ encodes the Euclidean geometry of
the line, plane,  three- and four-space, as well as the Minkowski geometry of the plane and four-space.

Finally, given a Clifford algebra in terms of a skew-field of quaternions, $\HH$, then
\begin{eqnarray}\label{1.1.6}
\HH= \HH_{F}
\end{eqnarray}
for some symplectic plane $\FF$. In fact, the tangent plane $\FF$, say, at any $j\in S^2 \subset \HH$ is symplectic
with symplectic structure $\omega$ given by
\begin{eqnarray}\label{lomega}
\omega(v,w)=<j\cdot v,w>=<j,v\cdot w> \quad \forall v,w\in \FF.
\end{eqnarray}
As easily seen, (\ref{1.1.5}) holds true. Thus the symplectic plane $\FF$ can be reconstructed from $\HH$; in this
sense $\HH$ is characteristic for $\FF$.

In fact the Clifford algebra $\HH$ in (\ref{1.1.6}), the skew-field of quaternions encodes all symplectic
planes in the three-dimensional real linear space $\EE$, as is easily seen.

As mentioned in section \ref{PrH} the Clifford algebra $\HH$ emanates from processes; since the symplectic space
$\FF$ can be reconstructed from its geometric Clifford algebras by (\ref{1.1.6}), similarly the symplectic space $\FF$ can also be reconstructed from process as well.

The construction of the geometric Clifford algebra of a \emph{2n-dimensional} symplectic space $\FF$ is based on
the following lemma (cf.~\cite{Greub_1975}):
\begin{lemma}\label{dec}
Let $\FF$ be a \emph{2n-dimensional} symplectic space with $\omega$ as (constant) symplectic form. The linear
space can be decomposed into the direct sum
\begin{eqnarray}\label{oplus}
 \FF=\oplus_s \FF_s
\end{eqnarray}
(where s runs from 1 to $n$) which admits a basis $\{e_1,...,e_{2n}\}$ such that the intersection
$\{e_1,...,e_{2n}\}\bigcap \FF_s$ is a symplectic basis of $\omega_s$, the restriction of $\omega$ to $\FF_s$ for
each $s$. In addition there is a unique scalar product $<,>_s$ for each s, such that $\omega_s$ is represented
uniquely by $J_s$ having (\ref{acs}) as its matrix and which satisfies (\ref{1.1.2}).

\end{lemma}

Hence there is a unique scalar product $<,>$ on $\FF=\oplus_s \FF_s$ yielding $<,>_s$ in lemma \ref{dec} if
restricted to $\FF_s$ for each s. From lemma \ref{splitF} each $\FF_s$ in (\ref{oplus}) is a phase space. If
$\HH_{F_s}$ denotes the Clifford algebra of $-<,>_s$ for all s, the Clifford algebra $\HH_F$ of $-<,>$ on $\FF$ is
given by the $2^{2n}$-dimensional algebra
\begin{eqnarray}\label{Pr}
\HH_{F}= {\otimes}_s \HH_{F_s}.
\end{eqnarray}
Here $s$ runs from $1$ to $n$ and ${\otimes}$ denotes the \emph{graded tensor product} (cf.~\cite{Greub_1978}.  This tensor product is
called the anticommutative tensor product) of the ${\ZZ}_2$ graded algebras $\HH_{F_s}$. Again here we will
call it the \emph{geometric Clifford algebra} of the symplectic space $F$; obviously $\HH_{F}$ is
\emph{characteristic} for $\FF$ . Clearly each $\HH_{F_s}$ is a subalgebra of $\HH_{F}$.

As an example, consider $m$ particles moving in $\RR^{3}$ endowed with the canonical scalar product and its
natural orthonormal basis. The configuration space is the m-fold direct sum $\RR^{3}\oplus....\oplus\RR^{3}$
which is isomorphic to the m-fold Cartesian product of $\RR^{3}$ with itself, a $3m$-dimensional linear space.
Let the $q$-coordinates be enumerated from $q_1$ to $q_{3m}$. The phase space is isomorphic to
$(\RR^{3}\oplus....\oplus\RR^{3})\times(\RR^{3}\oplus....\oplus\RR^{3})$, a $2\cdot3m$-dimensional linear space.
The first factor of this cartesian product constitutes all possible spatial positions of the $m$ particles,
while the second factor contains all the momenta. This phase space together with the canonical symplectic form
$\omega$, say, is a $2n$ dimensional symplectic space. Here $n=3m$. Now we regroup the coordinates in this phase
space. We take the plane $\FF_1$ generated by the first spatial coordinate $q_1$ and its moment $p_1$ together
with the canonical symplectic form. Passing to the second spatial coordinates and its momenta yields $\FF_2$ etc.
In this way we obtain
\begin{eqnarray}\label{ex}
\FF:=\oplus_s \FF_s\cong(\RR^{3}\oplus....\oplus\RR^{3})\times(\RR^{3}\oplus....\oplus\RR^{3}),
\end{eqnarray}
and endow it with the natural scalar product. Here s varies from $1$ to $3m$. The symplectic bases of $\FF$ in
which the canonical symplectic form $\omega$ decomposes according to lemma \ref{dec} is identical with the
canonical orthonormal basis of $\FF$ in (\ref{ex}).

\subsection{The Clifford Algebra determined by $sp(F)$}\label{sbsec3}

Another realization of the geometric Clifford algebra of the \emph{two-dimensional symplectic space} $\FF$ is
obtained by means of sp$(F)$, the Lie algebra of the symplectic group $Sp(F)$. At first we will construct a
natural splitting of sp$(F)$ and a symplectic structure on a two-dimensional subspace. The basis of this
construction is End\,$F$, the four-dimensional $\RR$-linear space of all ($\RR$-linear) endomorphisms of $\FF$. This space is split into
\begin{eqnarray}\label{Hgr}
\mbox{End} F=\RR \cdot id_{F}\oplus  sp(F)
\end{eqnarray}
where sp$(F)$ is the three-dimensional $\RR$-linear space of all \emph{traceless endomorphisms} of $\FF$. In
fact, this splitting is orthogonal with respect to the  natural scalar product $< , >_{End F}$, say, defined by
\begin{eqnarray}\label{trscalar}
<A,B>_{End F}:=\frac{1}{2}tr A\circ\widetilde B\qquad \forall A,B\in End {F}.
\end{eqnarray}
Here $\widetilde B$ is the adjoint of $B$. In fact the splitting (\ref{Hgr}) carries a natural \emph{Heisenberg
algebra structure} (cf.~\cite{Binz_Pods_2008}).

By means of this scalar product we observe that any almost complex structure on a two-dimensional linear
subspace of sp$(F)$ is a unit vector and vice versa any vector in the unit sphere $S^2 \subset \mbox{sp}(F)$ is an
almost complex structure on the tangent plane of $S^2$ at that vector (cf.~\ref{lomega}).

Moreover, let $\mu_{sp(F)}$ be a constant metric volume form, i.e. a determinant function
(cf.~\cite{Greub_1975}) on $sp(F)$, of the scalar product $<A,B>_{sp(F)}$ in (\ref{trscalar}).

For a given $J$ with matrix $\begin{pmatrix}\;\;0&1\\-1&0\end{pmatrix}$, the linear space sp$(F)$ decomposes into

$$
sp(F)=\RR\cdot J\oplus \Sigma
$$
where the matrix of an element in $\Sigma$ has the form $\begin{pmatrix}a&b\\b&-a\end{pmatrix}$, obviously a
\emph{symmetric matrix} with vanishing trace. (These matrices are formed with respect to an orthonormal
basis\index{orthonormal basis} in $F$). This yields
$$
\mbox{End} F=\RR\cdot id+\RR\cdot J+\Sigma
$$
and hence
$$
\Sigma=\mbox{span}\left\{a\cdot \left. \begin{pmatrix}1&0\\0&-1\end{pmatrix}, c \cdot
\begin{pmatrix}0&1\\1&0\end{pmatrix} \right| a,c \in \RR\right\}.
$$
(Thus sp\,$(F)$ is not closed under composition, because of (\ref{MTEND 2})). Also because of  (\ref{MTEND 2}), the
linear two-dimensional space $\Sigma$ inherits a \emph{symplectic structure} $\omega_\Sigma$ defined by
$$
\omega_\Sigma(A,B):=<\frac{1}{2}\cdot [A,B],J>_{End F}\qquad\forall A,B\in sp(F).
$$
Since $[A,B] \in \RR \cdot J$ for $A,B$ in $\Sigma$ with $\parallel A\parallel =\parallel B\parallel =1$
$$
\mu_{sp(F)}(A,B,J)=\omega_\Sigma(A,B),
$$
and hence
$$
\mu_{sp(F)}(A,B,J)=<\frac{1}{2}\cdot [A,B],J>_{End F}.
$$
This identifies $\frac{1}{2}\cdot [A,B]$ as the \emph{cross product} of $A,B\in\Sigma$ in $sp(F)$.

Therefore, by (\ref{quat}) the four-dimensional linear space End\,$F$ is turned into a skew-field $\HH_{sp(F)}$
isomorphic to the quaternions , i.e. isomorphic to the geometric Clifford algebra $\HH_F$ of the symplectic
space $\FF$. (Note End\,$F$ is a Clifford algebra with the composition as multiplication, as
mentioned in section \ref{CA}. It is not isomorphic to $\HH_{sp(F)}$).

We point out that $\Sigma$ and $\omega_\Sigma$ can be naturally identified with $\FF$ and $\omega$ respectively as
shown in (cf.~\cite{Binz_Pods_2008}). Hence
\begin{eqnarray}\label{HspF}
\HH_{F} \equiv \HH_{sp(F)}.
\end{eqnarray}
In the following we will use either picture.

\subsection{Poisson Algebras of the Phase Space}

In order to derive a characteristic Clifford algebra from the Poisson algebra of all \emph{homogenous
quadratic polynomials} on a two-dimensional phase space $\FF$, we will need to study the Poisson algebra of $\FF$.

Our first goal is to establish a natural isomorphism between the Poisson algebra ${\mathcal Q}$ of all
\emph{homogeneous quadratic polynomials} on the \emph{two-dimensional phase space} $\FF$
(cf.~\cite{Marsden_Ratiu_1994}) and $sp(\FF)$ ({cf.~\cite{Varadarajan_1984}). The key to this isomorphism will be
the construction of \emph{Hamiltonian vector fields} of homogenous quadratic polynomials
(cf.~\cite{Guillemin_Sternberg}).

The calculation in what follows can easily be visualized if we use the quaternions $\HH_{F}$ of the phase
space $\FF$. This skew-field contains $\FF$ as well as the line $\RR \cdot j$ which is perpendicular to $\FF$.

The \emph{coordinate system} in $\FF$ determined by the orthonormal unit vectors $e_q$ and $e_p:=j\cdot e_q$
respectively, of the (linear) \emph{coordinate functions}\index{coordinate function} $f_q$ and $f_p$ on $\FF$.  This
allows us to introduce the concept of a polynomial on $\FF$ in \emph{two variables} $q$ and $p$, say. The
collection ${\mathcal Q}$ of all homogenous quadratic polynomials on $\FF$ obviously forms an $\RR$-vector space.

Given a polynomial $pol$ on $\FF$, let $f_{pol}:\FF\too\RR$ be its polynomial function in the variables $q$ and
$p$. In the following we will identify the collection
\[
P_{\mathcal Q}:=\{f_{pol}|pol\in {\mathcal Q}\}
\]
with ${\mathcal Q}$ by identifying any  $pol$ with $f_{pol}$, if no confusion arises.

We begin our investigations of ${\mathcal Q}$ by the geometric study of the notion of Hamiltonian vector fields
on the two-dimensional phase space. Let $C^\infty(F,\RR)$ be the collection of all smooth $\RR$-valued functions
of $\FF$. It is an $\RR$-algebra under pointwisely defined operations. Given $f\in C^\infty(\FF,\RR)$ , its
\emph{Hamiltonian vector field} $X_f$ is defined by
\begin{eqnarray}\label{f9.8}
\omega(X_f, X)=df(X)
\end{eqnarray}
for any smooth vector field $X$ on $F$ (cf.~\cite{Marsden_Ratiu_1994}). Clearly, $X_f$ is smooth as well.

Throughout this paper, we will replace the vector field $X_f=({\bf a}_f, id)$ by its principal part ${\bf a}_f$,
mapping $\FF$ into itself. Thus (\ref{f9.8}) turns into
\begin{eqnarray}\label{f9.9}
\omega({\bf a}_f(h),k)=df(h)(k)\qquad\forall h,k\in \FF.
\end{eqnarray}
Here $df(h)(k)$ is the derivative of $f$ at $h$ evaluated at $k\in \FF$.

It follows from (\ref{1.1.3}) and (\ref{lomega}) that equation (\ref{f9.9}) can be rewritten in terms of the scalar product,
yielding
\[
\omega({\bf a}_f(h),k)=<{j\times \bf a}_f(h),k>=df(h)(k)\qquad\forall h,k\in \FF
\]
which by (\ref{1.1.5}) yields in turn a link to $\nabla f\in C^\infty(F,\RR)$, namely  that the
principal part ${\bf a}_f$ of the Hamiltonian vector field which is
\begin{eqnarray}\label{f9.10}
{\bf a}_f(h)=-j\times\nabla f(h)=-j\cdot \nabla f(h)\qquad\forall h\in \FF
\end{eqnarray}
where $\cdot$ denotes the multiplication in $\HH_{F}$.

A direct calculation in the orthonormal basis given by the unit vectors $e_{q}$ and $e_{p}$ shows:
\begin{eqnarray}\label{ff9.11}
{\bf a}_f(h)=\frac{\partial f(h)}{\partial p}\cdot e_{q}-\frac{\partial f(h)}{\partial q}\cdot e_{p}\qquad\forall
h \in \FF.
\end{eqnarray}

As an example, by (\ref{f9.10}) the Hamiltonian vector fields of the coordinate functions $f_q$ and $f_p$
obviously are
\begin{eqnarray}\label{f9.14}
{\bf a}_{f_q}=-j\cdot e_{q}=-e_{p}\qquad\text{ and }\qquad {\bf a}_{f_p}=-j\cdot e_{p}=e_{q}.
\end{eqnarray}
Clearly, $j=e_q\cdot e_p$ holds true.

We study the above notions by turning to $\mathcal Q\subset C^\infty(F,\RR)$. Of special interest
with respect to our goal is the linear dependence of ${\bf a}_f(h)$ on the position $h\in \FF$ for any $f\in
{\mathcal Q}$.  By (\ref{f9.9}) this means that
\begin{eqnarray}\label{Hamfeldislin}
{\bf a}_f \in \mbox{End} F;
\end{eqnarray}
hence both sides of (\ref{f9.9}) depend bilinearly on $h$ and $k$ for any $f\in {\mathcal Q}$.

The quality of  ${\bf a}_f$ will become clear if we notice that $\mathcal Q$ is generated by
$\frac{q^2}{2},\frac{p^2}{2}$ and $q\cdot p$.

 By (\ref{ff9.11}) the Hamiltonian vector fields of
$f_{\frac{q^2}{2}},f_{\frac{p^2}{2}}$ and $f_{q\cdot p}$ are the following for all $h \in F$ of the form
$h=q\cdot e_{q}+p\cdot e_{p}$:
\begin{eqnarray}\label{f9.21}
 {\bf a }_{f_{\frac{q^2}{2}}}(h)=-q\cdot e_{p} \qquad {\bf a }_{f_{\frac{p^2}{2}}}(h)=p\cdot e_{q}
\qquad  {\bf a }_{f_{q\cdot p}}(h)=q\cdot e_{q}-p\cdot e_{p}.
\end{eqnarray}

Their respective matrices $M({\bf a }_{f_{q^2}}),M({\bf a }_{f_{p^2}})$ and $M({\bf a }_{f_{q\cdot p}})$ formed
with respect to $e_{q}$ and $e_{p}$ (cf.~\ref{Hamfeldislin}) are
\begin{eqnarray} \label{9.31}
M({\bf a }_{f_{q^2}})=\begin{pmatrix}0&0\\-1&0\end{pmatrix}, \quad
M({\bf a }_{f_{p^2}})=\begin{pmatrix}0&1\\0&0\end{pmatrix}, \quad
M({\bf a }_{f_{q\cdot p}})=\begin{pmatrix}1&0\\0&-1\end{pmatrix}.
\end{eqnarray}
These matrices are all \emph{tracefree} which means they belong to the Lie algebra sp\,$(F)$.

Since the endomorphisms ${\bf a }_{f_{q^2}},{\bf a }_{f_{p^2}}$ and ${\bf a }_{f_{q\cdot p}}$ of $\FF$ generate
sp\,$(F)$, we have a surjective linear map
\begin{eqnarray}\label{ff9.18}
\mbox{ham}:{\mathcal Q}&\too& sp(F) \nonumber \\ \mbox{ham}\,(f)&:=&{\bf a }_{f}\qquad\forall f\in {\mathcal Q}.
\end{eqnarray}
ham is a \emph{linear isomorphism}\index{linear isomorphism} since both the domain and the range are
three-dimensional.

Next we will verify that ${\mathcal Q}$ forms a Poisson sub-algebra of $C^\infty(F,\CC)$. To this end, we define
a \emph{Poisson bracket} on $C^\infty(F,\RR)$  and show that $C^\infty(F,\RR)$ is a Poisson algebra
(cf.~\cite{deGosson_M}), which means that the $\RR$-linear space, $C^\infty(F,\RR)$,  has a natural Lie algebra
structure. Therefore, let $f,g\in C^\infty(F,\RR)$ with ${\bf a}_f$ and ${\bf a}_g$ being the principal parts of
their Hamiltonian vector fields. Obviously,
\[
\omega({\bf a}_f(h),{\bf a}_g(h))=df(h)({\bf a}_g(h)).
\]
Thus
\[
df({\bf a}_g)=<j,{\bf a}_f\times {\bf a}_g>.
\]
To evaluate the cross product\index{cross product} we use equation (\ref{ff9.11}) to obtain
\begin{eqnarray*}
{\bf a}_f\times {\bf a}_g& =& -\frac{\partial f}{\partial p}\frac{\partial g}{\partial q}\cdot e_{q}\times
e_{p}-\frac{\partial f}{\partial q} \cdot \frac{\partial g}{\partial p}\cdot e_{p}\times e_{q}
\end{eqnarray*}
and, therefore,
\begin{eqnarray}\label{f9.12}
{\bf a}_f\times {\bf a}_g=-\left(\frac{\partial f}{\partial q}\cdot\frac{\partial g}{\partial p}- \frac{\partial
f}{\partial p}\cdot \frac{\partial g}{\partial q}\right)\cdot j.
\end{eqnarray}

 Defining the \emph{Poisson bracket}\index{Poisson bracket} of
$f$ and $g$ in $C^\infty(F,\RR)$ by
\begin{eqnarray} \label{fgPoisson}
\{f,g\}:= \frac{\partial f}{\partial q}\cdot\frac{\partial g}{\partial p} - \frac{\partial f}{\partial p}\cdot
\frac{\partial g}{\partial q},
\end{eqnarray}
yields
\begin{eqnarray}\label{ff9.13}
\{f,g\}\cdot j=-{\bf a}_f\times {\bf a}_g;
\end{eqnarray}
hence this bracket can be expressed as
\[
\{f,g\}\cdot e=j\cdot ({\bf a}_f\times {\bf a}_g)\qquad\forall f,g\in C^{\infty}(F,\RR).
\]
They satisfy a Jacobi identity, i.e.~
\[
\{\{f_1,f_2\},f_3\}+\{\{f_2,f_3\},f_1\}+\{\{f_3,f_1\}f_2\}=0.
\]
Therefore, $C^\infty(F,\RR)$ is a Lie algebra\index{Lie algebra} under the Poisson bracket (\ref{fgPoisson})
(cf.~\cite{Varadarajan_1984}).

 Since $\{f,g\}\in C^\infty(F,\RR)$, we may investigate
${\bf a}_{\{f,g\}}$. Let $[{\bf a }_f,{\bf a }_g]$ denote the principal part of the \emph{Lie bracket}
\index{Lie bracket} $[X_f,X_g]$ defined by
\[
[{\bf a }_f,{\bf a }_g]:=d{\bf a }_g({\bf a }_f)-d{\bf a }_f({\bf a }_g).
\]
As is well known
\[
d\{f,g\}(h)=-\omega^a([{\bf a }_f,{\bf a }_g],h)\qquad\forall h\in \FF
\]
showing
\begin{eqnarray}\label{ff9.15}
{\bf a }_{\{f,g\}}=-[{\bf a }_f,{\bf a }_g]\qquad\forall f,g\in C^\infty(F,\RR).
\end{eqnarray}
Denoting by Ham\,$\FF$ the collection of all Hamiltonian vector fields on $\FF$, the linear isomorphism in
(\ref{ff9.18}) extends to the linear map
\begin{eqnarray}\label{ff9.21}
\mbox{ham}:C^\infty(F,\RR)\to \mbox{Ham}\, \FF
\end{eqnarray}
determined by
\[
\mbox{ham}\, f:={\bf a }_f\qquad\forall f\in C^\infty(F,\RR)
\]
which satisfies (\ref{ff9.15}).

Now focussing on ${\mathcal Q}$ again, the Poisson brackets\index{Poisson bracket}
$\{f_{\frac{q^2}{2}},f_{\frac{p^2}{2}}\},\{f_{q\cdot p}, f_{\frac{q^2}{2}}\}$ and $\{f_{q\cdot
p},f_{\frac{p^2}{2}}\}$ are
\begin{eqnarray}\label{f9.22}
\{f_{\frac{q^2}{2}},f_{\frac{p^2}{2}}\}=f_{q\cdot p} \quad \{f_{q\cdot
p},f_{\frac{q^2}{2}}\}=f_{q^2}\quad\{f_{q\cdot p},f_{\frac{p^2}{2}}\}=f_{p^2}
\end{eqnarray}
which are easily calculated either from  (\ref{fgPoisson}) or (\ref{ff9.13}). Hence we have shown
(cf.~\cite{Guillemin_Sternberg})
\begin{proposition}
${\mathcal Q}$  closes under the Poisson brackets and is a Poisson sub-algebra of $C^\infty(\FF,\RR)$.
\end{proposition}
The formalism we have used here to introduce Hamiltonian vector fields and Poisson brackets is adapted to our
setting for two-dimensional phase spaces. For a general formalism we refer to \cite{Marsden_Ratiu_1994} or
\cite{Guillemin_Sternberg}.

\subsection{Clifford Algebra Structure based on ${\mathcal Q}$.}\label{gnPoisson}

Next, we extend the Poisson algebra of all homogenous quadratic polynomials on a \emph{two-dimensional} phase
space to a skew-field of quaternions, a Clifford algebra and generalize this to \emph{2n-dimensional} symplectic
space.

Because of (\ref{HspF}) we will now relate the Lie algebra sl$(F)$ of SL$(F)$, the special linear group
of $\FF$, to ${\mathcal Q}$ (cf.~subsection \ref{sbsec3}). From (\ref{Hamfeldislin}) we have
\begin{eqnarray}
\mbox{sp}(F)=\mbox{sl}(F).
\end{eqnarray}


According to (\ref{ff9.11}), for any $f\in{\mathcal Q}$ the trace free
matrix $M({\bf a}_f)$ of ${\bf a}_f$ is
\begin{eqnarray}\label{f9.16}
M({\bf a}_f)=\begin{pmatrix}\frac{\partial f}{\partial p}(e_{q})&\frac{\partial f  }{\partial p}(e_{p})\\
-\frac{\partial f  }{\partial q}(e_{q})&-\frac{\partial f  }{\partial q}(e_{p})\end{pmatrix}.
\end{eqnarray}
Moreover, a direct calculation provides us with the key result on $ham$ in (\ref{ff9.18}) for the construction
of the Clifford algebra containing $\mathcal Q$:

\begin{proposition}\label{hamLiealgebraisom}
\begin{eqnarray}
ham:{\mathcal Q}&\too& sp(F)
\end{eqnarray} is a Lie algebra isomorphism satisfying
\begin{eqnarray}\label{f8.34}
ham\{f,g\}=[{\bf a}_f,{\bf a}_g]_{sl(F^a)}\qquad\forall f,g\in {\mathcal Q}.
\end{eqnarray}
Here
\[
[{\bf a}_f,{\bf a}_g]_{sl(F^a)}:={\bf a}_f\circ {\bf a}_g-{\bf a}_g\circ{\bf a}_f\qquad\forall f,g\in Q
\]
is the commutator in sl$(F^a)$.
\end{proposition}

Now we enlarge $\mathcal Q$ to $\HH_{\mathcal Q}:=\mathcal Q \oplus\RR\cdot e$ and extend $ham$ in proposition
\ref{hamLiealgebraisom} to
\begin{eqnarray}\label{hamQ}
\mbox{ham}:{\HH_{\mathcal Q}}&\too& \mbox{End} F
\end{eqnarray}
by sending $e$ to the identity $id$. This is a linear isomorphism. As shown in section 2.2 the linear space
End\,$F$ carries a natural Clifford algebra structure arising from the Lie algebra sp$(F)$. This structure can
be pulled back to $\HH_{\mathcal Q}$ by $ ham$. The resulting skew-field is called here the \emph{Poisson
Clifford algebra} of the phase space and is also denoted by $\HH_{\mathcal Q}$. Compared with the defining
equation (\ref{quat}) for the product in $\HH_F$, the Poisson bracket on ${\mathcal Q}$ serves as the cross
product. By construction we have
\begin{eqnarray}
\HH_\FF=\HH_{sp(F)}\cong\HH_{\mathcal Q}
\end{eqnarray}
as Clifford algebras.

In the case where the phase space $\FF$ is \emph{2n-dimensional}, we proceed as in section \ref{sbsec1}. We decompose $\FF$
into $\FF=\oplus_s \FF_s$ and form for each $s$ the Clifford algebra $\HH_{\mathcal Q_s}$.

Hence $\otimes_s{\HH_{\mathcal Q_s}}$ is a Clifford algebra, called here the \emph{Poisson Clifford algebra} of
the phase space $F$. Obviously
\begin{eqnarray}
\HH_F={\otimes_s}\HH_s\cong {\otimes}_s{\HH_{\mathcal Q_s}}
\end{eqnarray}
holds true.

$\mathcal Q$, the collection of all functions of homogenous quadratic polynomials of $\FF$, is the image of a
natural subspace of the Poisson Clifford algebra $\otimes_s \HH_s$, as seen as follows: Based on (\ref{f9.21})
we restrict $ham$ in (\ref{ff9.21}) to $\mathcal Q$ and calculate in ${\otimes_s}\HH_{sp(F_s)}$. The tensor
$q_s\cdot p_re\otimes...\otimes e_s\otimes...\otimes e_r...\otimes e$ in ${\otimes_s}\HH_{sp(F_s)}$ applied to
an element of $\FF=\oplus_s \FF_s$ in (\ref{oplus}) (via the canonical scalar product) yields $q_s \cdot p_r$ for
all $s,r \leq n$. Thus both $q_s\cdot p_re\otimes...\otimes e_s\otimes...\otimes e_r...\otimes e$ as well as
$q_s\cdot p_re\otimes...\otimes e_r\otimes...\otimes e_s...\otimes e$ map linearly to $q_s \cdot p_r$ in
$\mathcal Q$ for all $s,r \leq n$. In this way we obtain all of $\mathcal Q$.

\subsection{A realization of $\HH_F$ consisting of Operators}\label{Quantm}

We close our studies of characteristic Clifford algebras of symplectic spaces by  introducing for any
finite-dimensional phase space $\FF$ a realization of $\HH_F$ in terms of a Clifford algebra consisting of
\emph{Hermitian operators}. This will allow us to associate eigen-values to any element of $\HH_F$. (For an
alternative approach, consult \cite{bjh_2011process}).

This realization is constructed first for a two-dimensional phase space $\FF$ and then extended to
$2n$-dimensions  $\FF=\oplus_s \FF_s $ by means of (\ref{Pr}).

Given a two-dimensional phase space $\FF$, this construction is based on the \emph{quantization map}
$$
Q:{\mathcal Q}\too \mbox{End}(L^2(\RR,\CC))
$$
(cf.~\cite{Binz_Pods_2008}). It is defined to be the composition of
\begin{eqnarray}
\mathcal Q\stackrel{ham}{\too} sp(F)\overset{-i \cdot dU_{Mp}}{\too}\mbox{End}(L^2(\RR,\CC)),
\end{eqnarray}
i.e.
\begin{eqnarray}\label{quantmap}
\ Q:=-i\cdot dU_{Mp}\circ \mbox{ham}.
\end{eqnarray}

The right hand composite is made up by the (injective) infinitesimal metaplectic representation $dU_{Mp}$
(cf.~\cite{Binz_Pods_2008} or~\cite{deGosson_M}) multiplied by the factor $-i$ and the Lie algebra isomorphism
$ham$ in (\ref{ff9.18}). The values of $dU_{Mp}$ are \emph{skew-symmetric operators} acting on the Hilbert
space $L^2(\RR,\CC)$, hence $\mp i \cdot dU_{Mp}$ maps into a collection of Hermitian operators. The choice of the
factor $-i$ in $Q$ (cf.~(\ref{quantmap})) was made to guarantee that
\begin{eqnarray}\label{Poisoncom}
Q(\{f,g\})=[Q(f),Q(g)]
\end{eqnarray}
holds true for any pair of polynomial function $f,g\in {\mathcal Q}$. Thus $Q$ \emph{quantizes} any homogenous
quadratic polynomial defined on $\FF$.
(The linear isomorphism $ham$ in (\ref{ff9.18}) can be defined for any finite-dimensional phase space $\FF$
(cf.~\cite{Guillemin_Sternberg})).

The isomorphism (\ref{quantmap}) obviously extends to a Clifford algebra isomorphism also denoted by $ham$. This
is to say we have

\begin{eqnarray}\label{hamq}
\mbox{ham}:\HH_{\mathcal Q}&\too&\HH_{sp(F)},
\end{eqnarray}
mapping $e \in \HH_{\mathcal Q}$ to $id$.

We extend the monomorphism $dU_{Mp}$ to all  End $F=\RR\cdot id_F\oplus sp(F)$ as a linear injective map
(still called $dU_{Mp}$)
\[
dU_{Mp}:{\RR\cdot id_F\oplus sp(F)}\too{\mbox{End}(L^2(\RR,\CC))}
\]
by setting
\[
dU_{Mp }(id_F)= id_{\mbox{End}(L^2(\RR,\CC)}.
\]
Together with (\ref{quantmap}), this means we have a linear map
\[
\ Q:\HH_F\too{\mbox{im}\; Q}.
\]
Here im\,$Q$ is the (four-dimensional) image of $Q$.

Pushing forward by $Q$, the skew-field structure of $\HH_F$ to im\,$Q$ yields the structure of a skew-field of
quaternions on im\,$Q$, a Clifford algebra as mentioned above. It is denoted by $\HH_Q$.

To extend the quantization map $Q$ to $\HH_F={\otimes_s}\HH_s$ for an \emph{$2n$-dimensional} phase space $\FF$ as
in ({\ref{Pr}), we form the quantization map for each $s$ yielding $Q_s$ of which the image is im\,$Q_s$ and set
\begin{eqnarray}
Q:={\otimes_s}Q_s.
\end{eqnarray}
The image of $Q$ is the graded tensor product
$$
{\otimes_s}{\mbox{im}\,Q_s},
$$
a finite dimensional Clifford algebra, referd to as  the \emph{realization} of  $\HH_F$ by means of
\emph{Hermitian operators}.

\section{Acknowledgement}
We are deeply indebted to the generosity of Georg Wikman in sponsoring the Askloster Seminars where different aspects of this work were presented and discussed.  Without those meetings the work presented in this paper would not have been possible.

\bibliographystyle{plain}
\bibliography{Bib_Binz_Hiley}

\end{document}